\documentclass[12pt]{iopart}
\usepackage{iopams} 
\usepackage{epsfig}
\begin{document}
\title{Perspectives for future neutrino oscillation experiments with
accelerators: beams, detectors and physics}

\author{Pasquale Migliozzi\dag}

\address{\dag\ INFN, Sezione Napoli, Italy}

\ead{pasquale.migliozzi@cern.ch}

\begin{abstract}
In recent years great progress toward the understanding of the mixing
in the leptonic sector has been made.  Nonetheless, this field of
research is just at the beginning. Further advance by accelerator
based neutrino oscillation experiments requires new beams and
detectors to reach the wanted physics goals. In this paper we review
the next possible steps that can be done for neutrino oscillation
experiments with accelerators.
\end{abstract}

%\submitto{\JPG}
%\pacs{1315, 9440T}

%\maketitle

%
%%%%%%%%%%%%%%%%%%%%%%%%%%%%%%%%%%%%%%%%%%%%%%%%%%%%%%%%%%%%%%%%%%%%%%%%%%%%%%%
% Introduction
%%%%%%%%%%%%%%%%%%%%%%%%%%%%%%%%%%%%%%%%%%%%%%%%%%%%%%%%%%%%%%%%%%%%%%%%%%%%%%%
%

\section{A Brief Introduction to the Present Status of Neutrino Oscillation Searches}

The hypothesis of neutrino oscillations~\cite{Pontecorvo:1957yb} is
strongly supported by atmospheric~\cite{atmo}, solar~\cite{solar},
accelerator~\cite{k2k} and reactor~\cite{KAMLAND} neutrino data. If we
do not consider the claimed evidence for oscillations by the LSND
experiment~\cite{Athanassopoulos:1998pv}, that must be confirmed or
excluded by the ongoing MiniBooNE experiment~\cite{miniboone},
oscillations in the leptonic sector can be accommodated in the three
family Pontecorvo-Maki-Nakagawa-Sakata (PMNS) mixing matrix $U_{PMNS}$

\begin{eqnarray}
\label{pmns}
 U_{PMNS} =
\left (
\begin{array}{ccc}
c_{12}c_{13} &        s_{12}c_{13} &       s_{13}e^{-i\delta} \\
-s_{12}c_{23}-c_{12}s_{23}s_{13}e^{i\delta} &  c_{12}c_{23}-s_{12}s_{23}s_{13}e^{i\delta} &  s_{23}c_{13} \\
s_{12}s_{23}-c_{12}c_{23}s_{13}e^{i\delta} & -c_{12}s_{23}-s_{12}c_{23}s_{13}e^{i\delta} &  c_{23}c_{13}
\end{array}
\right )
\end{eqnarray}

where the short-form notation $ s_{ij} \equiv \sin \theta_{ij}, c_{ij}
\equiv \cos \theta_{ij}$ is used.  Further Majorana phases have not
been introduced, since oscillation experiments are only sensitive to
the two neutrino mass squared differences $\Delta m^2_{12}, \Delta
m^2_{23}$ and to the four parameters in the mixing matrix of
eq.~(\ref{pmns}): three angles and the Dirac CP-violating phase,
$\delta$.

There are several global fits of all available data. As an example we
report in Tab.~\ref{tab:summary} the results taken from
Ref.~\cite{Maltoni:2003da}. Notice that there are no direct
measurements of the mixing angle $\theta_{13}$, but only the upper
limit given by the CHOOZ reactor
experiment~\cite{Apollonio:2002gd}. The global analysis in
Ref.~\cite{Maltoni:2003da} has given a bound on $\theta_{13}$:
$\sin^2\theta_{13} \leq 0.03$ at 90\% C.L..

\begin{table}[tbp]
\begin{center}
{\small
\begin{tabular}{|c|c|c|c|c|}
\hline parameter & best fit & 2$\sigma$ & 3$\sigma$ & 5$\sigma$ \\
\hline $\Delta m^2_{21}\: [10^{-5}\mbox{eV}^2]$ & 6.9 & 6.0--8.4 &
5.4--9.5 & 2.1--28\\ \hline $\Delta m^2_{31}\: [10^{-3}\mbox{eV}^2]$ &
2.6 & 1.8--3.3 & 1.4--3.7 & 0.77--4.8\\ \hline $\sin^2\theta_{12}$ &
0.30 & 0.25--0.36 & 0.23--0.39 & 0.17--0.48 \\ \hline
$\sin^2\theta_{23}$ & 0.52 & 0.36--0.67 & 0.31--0.72 & 0.22--0.81 \\
\hline
\end{tabular}
}
\end{center}
\caption{ Best-fit values, 2$\sigma$, 3$\sigma$ and 5$\sigma$
      intervals (1 d.o.f.) for the three-flavor neutrino oscillation
      parameters from global data including solar, atmospheric,
      reactor (KamLAND and CHOOZ) and accelerator (K2K) experiments.}
\label{tab:summary}

\end{table}

The next steps on the way of a full understanding
of neutrino oscillations by using neutrino beams produced at
accelerators are

\begin{itemize}
\item confirm the source of atmospheric neutrino oscillations,
i. e. observe the oscillation $\nu_\mu\rightarrow\nu_\tau$;
\item measure the remaining parameters of the PMNS mixing matrix:
$\theta_{13}$ and $\delta$;
\item measure the sign of $\Delta m^2_{23}$;
\item perform precision measurements of the angles
$\theta_{12}$ and $\theta_{23}$, and of
$\Delta m^2_{12}$ and $\Delta m^2_{23}$.
\end{itemize}

It is worth noting that there are other searches (like $\beta$-decay
and double-$\beta$ decay experiments, and space experiments studying
anisotropies in cosmic background radiation) which provide very
important information like the absolute value of the neutrino mass or
whether the neutrino is a Dirac or a Majorana particle. For a
comprehensive review of these experiments we refer to~\cite{review}.

\section{Neutrino Beams}

Current neutrino oscillation experiments are based on beams where
neutrinos come from the decay of mesons produced in the interaction of
high energy protons impinging onto a target (typically Be or
graphite). However, such conventional beams have some limitations (see
Section~\ref{conventional}) that could be overcome by using new
beam-line concepts: $\beta$-beams (Section~\ref{betabeam}) or Neutrino
Factories (Section~\ref{nufact}). For a comprehensive discussion of
future beams and their comparison we refer to~\cite{Apollonio:2002en}
and references therein.

\subsection{Conventional Neutrino Beams}
\label{conventional}

One can identify the main components of a conventional neutrino beam
line at a high energy accelerator as

\begin{itemize}
\item the target onto which protons are sent to produce pions and kaons;
\item the focusing system which guides the mesons
  along the desired neutrino beam direction;
\item the decay tunnel (usually evacuated) where mesons decay
  and produce neutrinos and muons.
\end{itemize}

From meson decay kinematics it follows that the neutrino energy is given by

\begin{equation}
E_\nu = {\frac{m^2_{\pi(K)}-m^2_\mu}{m^2_{\pi(K)}}}\frac{E_{\pi(K)}}{(1+\gamma^2
\theta^2)}
\label{enu_pi_K}
\end{equation}

where $\gamma$ is the Lorentz boost of the parent meson, $E_{\pi(K)}$
its energy and $\theta$ the angle of the neutrino with respect to the
meson flight direction.

There are three types of conventional neutrino beams: the Wide Band
Beams (WBB), the Narrow Band Beams (NBB) and the Off-Axis Beams
(OAB). WBB are characterized by a wide energy spectrum (they could
spread over a couple of order of magnitude) and correspondingly high
neutrino flux. Given these features, WBB are the optimal solution to
make discoveries. The drawback is that, if the signal comes from a
small part of the energy spectrum, it could be overwhelmed by the
background also induced by neutrinos outside the signal
region. Conversely, NBB may produce almost monochromatic energy
spectra. This can be obtained by selecting a small momentum bite of
the parent $\pi$ and $K$. However, the neutrino yield is significantly
reduced. This is an important drawback for oscillation searches.

A good compromise between the requirements of a high flux and a narrow
energy spectrum is obtained by means of Off-Axis
Beams~\cite{beavis}. This technique involves designing a beam-line
which can produce and focus a wide range of mesons in a given
direction (as in the WBB case), but then putting the detectors at an
angle with respect to that direction. Since the pion decay is a
two-body decay, a given angle between the pion direction and the
detector location corresponds to a given neutrino energy (almost)
independently of the pion energy. Furthermore, the smaller fraction of
high energy tails reduces the background from neutral-current (NC)
events, which can be mistaken for a $\nu_e$ charged-current (CC)
interaction due to the early showering of gamma's from the $\pi^0$
decay.

It is worth noting that, independent of the adopted solution, there
are common problems to all conventional neutrino beams

\begin{itemize}
\item the hadron yield in the proton-target interaction has large
uncertainties due to lack of data and to theoretical difficulties in
describing hadronic processes. This implies difficulties in predicting
the neutrino flux and spectrum with good accuracy;
\item in addition to the dominant flavor in the beam (typically
$\nu_\mu$) there is a contamination (at the few percent level) from
other flavors ($\bar{\nu}_\mu\,,\nu_e~\mbox{and}~\bar{\nu}_e $).
\end{itemize}

The knowledge of the beam spectrum and composition has a strong
impact both on the precision measurements of the angle $\theta_{23}$,
on the mass squared difference $\Delta m^2_{23}$ and on the
sensitivity to the mixing angle $\theta_{13}$. For instance, from the
CHOOZ limit on $\theta_{13}$ we know that the
$\nu_\mu\rightarrow\nu_e$ appearance probability is smaller than 5\%,
which is of the same order of magnitude of the beam
contamination. Therefore, the observation of $\nu_e$ appearance and
the related $\theta_{13}$ measurement are experimentally hard. The
experimental problem related to the knowledge of the beam, namely the
usage of a close detector, is addressed later in this Section.

In the last years a new concept of conventional beam (the so-called
``Super-Beam'') has been put forward in order to maximize the
sensitivity to $\theta_{13}$. Super-Beams will provide a much higher
neutrino flux, but at low energy (below 1 GeV). This will open the
possibility to perform long-baseline experiments with high statistics
and tuned at the oscillation maximum even at moderate distances
between source and detector.

In order to improve the knowledge of the beam, conventional beam-lines
are usually equipped with close detectors whose aim is to help in the
prediction of the flux and spectrum of the neutrino beam in absence of
oscillations (close to far extrapolation). In addition, they measure
the intrinsic contamination (in particular the
$\nu_e~\mbox{and}~\bar{\nu}_e $ components), the background rejection
capabilities of the detector and the neutrino cross-sections with high
statistics.

%\begin{figure}
%\begin{center}
%\includegraphics[width=3.in]{myoa2deg-spectrum.ps}
%\end{center}
%\caption{Expected neutrino spectra at different locations~\cite{kaji}.}
%\label{jhfspectrum}
%\end{figure}

\begin{figure}
\begin{center}
\includegraphics[width=2.in]{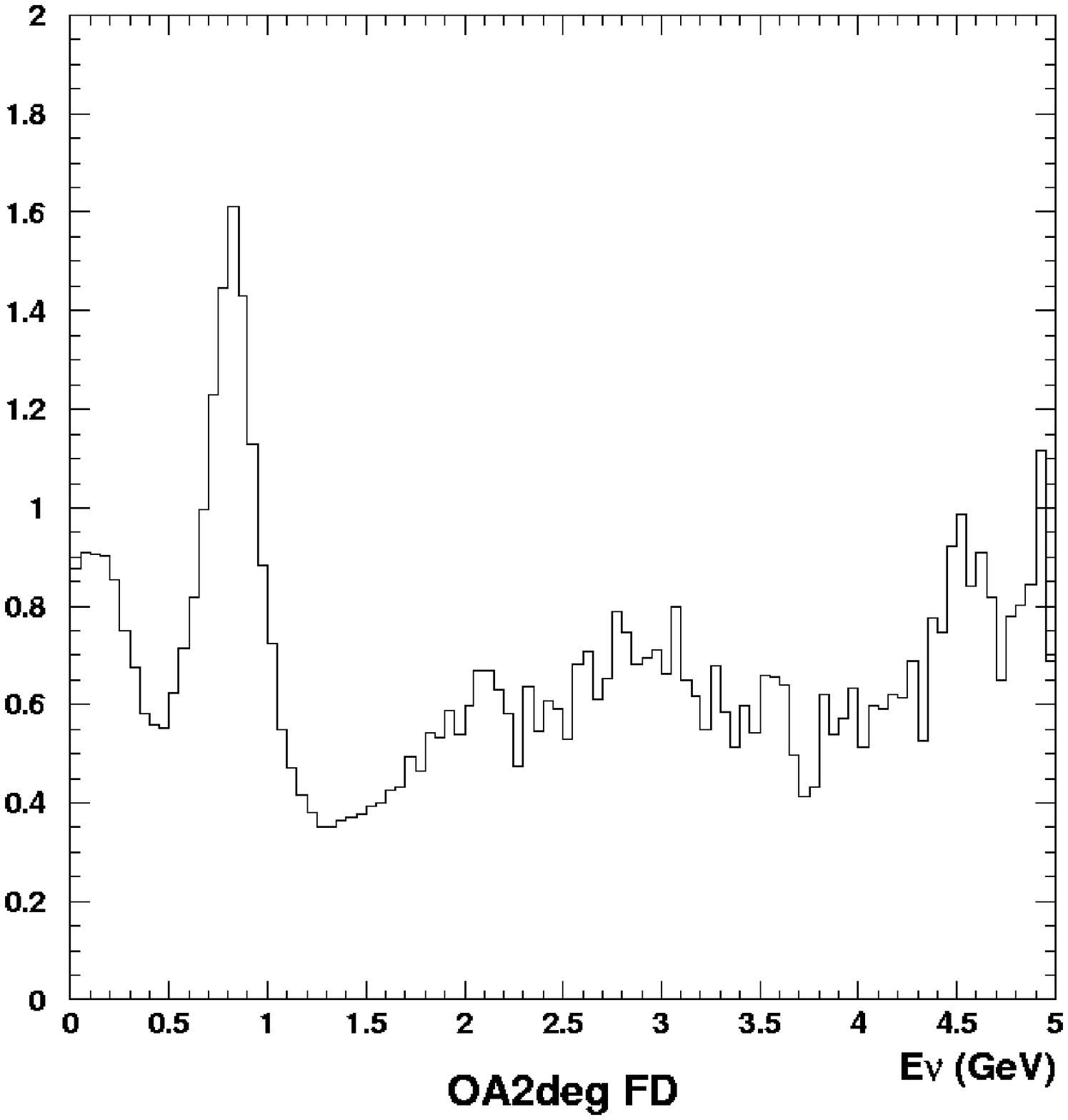} 
\includegraphics[width=2.in]{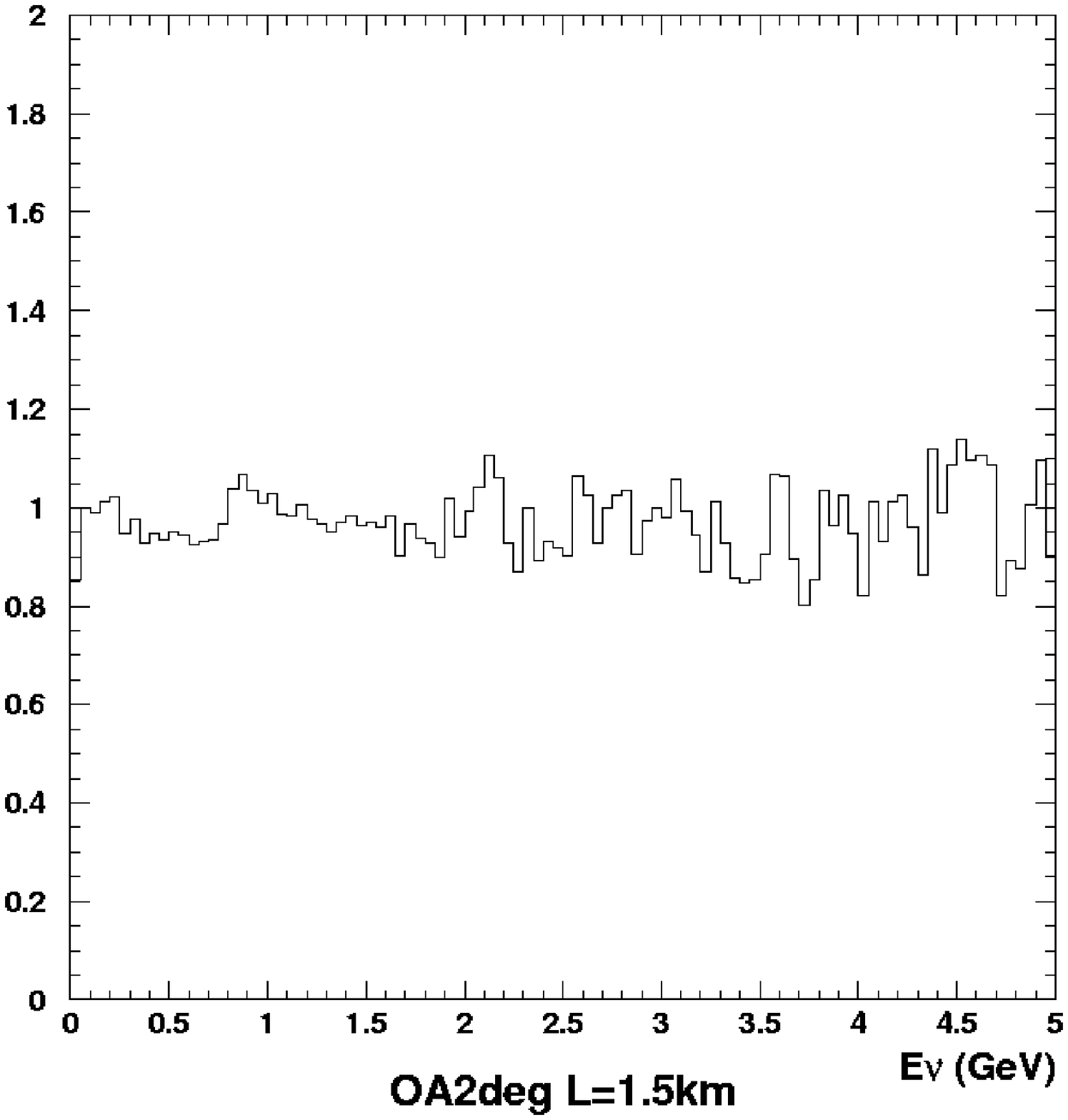}
\end{center}
\caption{Far to close ratio as function of the close detector distance
from the production target~\cite{kaji}: left panel 0.28~km, right
panel 1.5~km.}
\label{jhf2km}
\end{figure}

Recently the J-PARC Collaboration pointed out, profiting of the
experience gathered within the K2K Collaboration, that a single close
detector is not sufficient to achieve systematic errors on the
knowledge of the beam at the far detector better than
$\sim10\%$. Therefore, they proposed to build two close detectors: one
located at 250~m from the production target and a second at about
2~km. Indeed, it turned out that at 2~km from the production target,
the spectrum becomes almost identical to the one for Super-Kamiokande
at 295~km distance. The far to close
ratio as a function of the close detector distance is shown in
Fig.~\ref{jhf2km}. The ratio of the spectra at 295~km and 2~km deviate
from unity only about 5\%~\cite{Kobayashi:sc}. The small spectrum
correction enables high precision prediction of far spectrum from the
observation at the close detector. Consequently, it is possible to
perform high precision measurements of the oscillation parameters.

\subsection{$\beta$-beams}
\label{betabeam}

A $\beta$-beam~\cite{Zucchelli:sa} is made by accelerating radioactive
ions with a short beta-decay lifetime, by storing them in a ring with
straight sections and by letting them decay. The focusing of the beam
is provided by the Lorentz boost. Having the possibility to accelerate
either $\beta^-$ (e.g. $^6$He) or $\beta^+$ (e.g. $^{18}$Ne) ions,
pure $\bar{\nu}_e$ or pure ${\nu}_e$ beams can be produced,
respectively. In order to illustrate the value of the $\beta$-beam
concept, we briefly discuss the production of an anti-neutrino beam. A
good beta-emitter for anti-neutrino production is the
$^6\mbox{He}^{++}$ ion that decays into
$^6_3\mbox{Li}^{++}e^-\bar{\nu}_e$ with a $\beta$-decay endpoint
($E_0$) of about 3.5~MeV. The anti-neutrino spectrum is precisely
known from laboratory measurements of the associated electron, since
$E_e + E_\nu \approx E_0$. Since the ion is spin-less, decays at rest
are isotropic. When ions are accelerated ($\gamma$ values up to 150
are possible) the neutrino transverse momentum in the laboratory frame
is identical to that observed in the rest frame, while the
longitudinal momentum is multiplied by a factor $\gamma$. Therefore,
neutrino beam divergence is of the order of $1/\gamma$ (less than
10~mrad for $\gamma=100$), and the average neutrino energy in the
forward direction is $2\gamma\mbox{E}_{cms}\sim$500~MeV.

The technical feasibility of accelerating ions, although at relatively
low energies, has been already demonstrated in nuclear physics
experiments such as at ISOLDE at CERN. Given the small neutrino
energy, a potential drawback of this approach is the substantial
background from atmospheric neutrinos. To overcome this problem, ion
beams should be bunched. At present, this is a major technical issue.

Summarizing, the main features of a neutrino beam based on the
$\beta$-beam concept are

\begin{itemize}
\item the neutrino beam energy is low and neutrinos are well
focused (particularly important for long-baseline experiments);
\item the beam energy depends on the $\gamma$ factor. The ion
accelerator can be tuned to optimize the sensitivity of the
experiment;
\item the neutrino beam contains a single flavor with an energy
spectrum and intensity known a priori. Therefore, unlike conventional
neutrino beams, close detectors are not necessary;
\item neutrino and anti-neutrino beams can be produced with a
comparable flux.
\end{itemize}

\subsection{Neutrino Factory}
\label{nufact}

The first stage of a Neutrino Factory is similar to that of a
Super-Beam. Namely, protons are sent onto a target producing pions and
kaons that are collected by means of magnetic lenses. However, while
in those beams hadrons are let decay launching neutrinos toward the
detector site, in a Neutrino Factory daughter muons are collected and
accelerated in a ring with long straight sections. Muon decays in each
straight section generate highly collimated neutrino beams. If $\mu^+$
are stored, $\mu^+\rightarrow e^+\nu_e\bar{\nu}_\mu$ decays generate a
beam consisting of 50\% $\nu_e$ and 50\% $\bar{\nu}_\mu$. Similarly,
if $\mu^-$ are stored the beam consists of 50\% $\nu_\mu$ and 50\%
$\bar{\nu}_e$. Since the kinematics of muon decay is well known, we
expect minimal systematic uncertainties on the neutrino flux and
spectrum. Hence, compared to conventional neutrino beams, Neutrino
Factories provide $\nu_e$ and $\bar{\nu}_\mu$ beams or $\nu_\mu$ and
$\bar{\nu}_e$ beams, with small systematic uncertainties on the flux
and spectrum. Radiative effects on the muon decay have been calculated
and amount to about $4\times10^{-3}$ with a much smaller
error. Overall, the flux is expected to be known with a precision of
the order of $10^{-3}$. Another important feature of a Neutrino
Factory beam is its sharp cut-off at the energy of the stored
muons. In a conventional neutrino beam there is a high-energy tail
which, as already mentioned, gives rise to background from NC events
in which a leading $\pi^0$ is misinterpreted as an electron, faking
$\nu_\mu\rightarrow\nu_e$ signal.  Furthermore, the possibility to
store high-energy muons that in turn produce high-energy neutrinos
opens the study of oscillation channels like
$\nu_\mu\rightarrow\nu_\tau$ and $\nu_e\rightarrow\nu_\tau$, whose
combined physics potential has been discussed in~\cite{donini}.

\subsection{Beam-line Summary}

\begin{figure}
\begin{center}
\includegraphics[width=4.in]{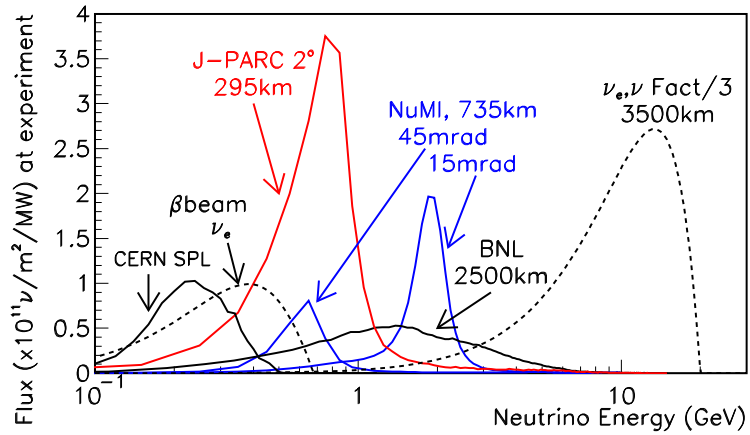}
\end{center}
\caption{Neutrino fluxes for some of the beam-lines discussed in the
previous Sections (from Ref.~\cite{Harris:2003rc})}
\label{beam_summ}
\end{figure}

The expected fluxes for some of the future proposed beam-lines are
shown in Fig.~\ref{beam_summ}~\cite{Harris:2003rc}. The fluxes are
normalized to 1~MW proton power, except for the $\beta$-beam, which
uses considerably less proton power. Notice that the fluxes are given
at the distance where the detector is expected to be located.

\section{Detector Technologies}

\subsection{Water Cerenkov Detector}
After the observations of solar and supernovae neutrinos, the
 discovery of neutrino oscillations in the atmospheric sector
 represents another great success of the Cerenkov technique. The main
 advantages of water Cerenkov detectors are both economical and
 scientific. The target material and the possibility to instrument
 only the surface of the target make these detectors relatively
 cheap. The charged leptons are identified through the detection of
 Cerenkov light. The features of the ring are used for particle
 identification. A muon crossing the detector scatters very
 little. Therefore, the associated ring has very sharp
 edges. Conversely, an electron scatters (showers) much more,
 producing rings with ``fuzzy'' edges. The efficiency to identify an
 electron or a muon is larger than 99\% for both leptons. The total
 measured light gives an estimate of the lepton energy, while the time
 measurement provided by each photomultiplier determines the outgoing
 lepton direction. By combining all this information it is possible to
 fully reconstruct the energy, the direction and the flavor of the
 incoming neutrino. It is worth noting that the procedure discussed
 above is suitable only for quasi-elastic events ($\nu_l n \rightarrow
 l^- p$). Indeed, for non quasi-elastic events there are other
 particles in the final state, carrying a large energy fraction, that
 either are below the Cerenkov threshold or are neutrals. Therefore,
 the total neutrino energy is not very well measured. Furthermore, the
 presence of more than one particle above threshold produces more than
 one ring, spoiling the particle identification capability of the
 detector.

Recently, a Cerenkov detector using heavy water as target became
operational (SNO). Besides the features discussed above, the SNO
detector is also able to identify NC neutrino interactions (see
Ref~\cite{Boger:1999bb} for details) through the detection of the
neutron produced in the reaction $\nu_l d\rightarrow \nu_l p n
$. Thanks to the SNO results on the measurement of NC interactions of
neutrinos coming from the Sun, it has been possible to clarify the
longstanding solar neutrino problem. Namely, it has been proved that
electron neutrinos produced in the Sun oscillate into active neutrinos
(muon or tau).

Given its cost effectiveness and its excellent performance at low
neutrino energies, the Cerenkov technique is often considered to
operate at the future neutrino beam-lines providing low energy beams.

%\begin{figure}
%\begin{center}
%\includegraphics[width=2in]{./ev_18_mu.eps}
%            \includegraphics[width=2in]{./ev_21_e.eps}
%\end{center}
%    \caption{Three-dimensional views of events induced by
%atmospheric neutrinos in the Super-Kamiokande detector. Left panel :
%muon ring induced by a $\nu_\mu$CC interaction. Right panel: electron
%ring produced by a $\nu_e$.}
%    \label{skevents}
%\end{figure}

\subsection{Magnetized Iron Calorimeter}
\label{irondet}

Magnetized iron calorimeters are used since the eighties. The
MINOS Collaboration has built a magnetized iron calorimeter to study
neutrino oscillations at the atmospheric scale by using the NuMI long-baseline
beam~\cite{minos}. We use the MINOS performance as a reference for this
kind of technology. The detector is composed of 2.54~cm-thick steel
planes interspersed with planes of 1~cm-thick and 4.1~cm-wide
scintillator strips. The iron is magnetized to an average field of
about 1.5~T. Simulations, as well as test beam results, show that the
energy resolution of this tracking calorimeter is
$55\%/\sqrt{E\mbox{(GeV)}}$ and $23\%/\sqrt{E\mbox{(GeV)}}$ for
hadronic and electromagnetic showers, respectively. This technology is
particularly suited for the reconstruction $\nu_\mu$CC events,
while the electron identification is rather poor. Therefore,
magnetized iron calorimeters are planned to be used to study either
$\nu_\mu$ appearance in a pure $\nu_e$ beam or $\nu_\mu$ disappearance
in a well known $\nu_\mu$ beam. It is worth noting that the presence
of the magnetic field is essential in order to tag the (anti-)neutrino
in the final state. This technology has been proposed to study the
so-called ``golden channels'' $\nu_e\rightarrow\nu_\mu$ and
$\bar{\nu}_e\rightarrow\bar{\nu}_\mu$~\cite{Cervera:2000kp} at a
Neutrino Factory.

\subsection{Hybrid Emulsion Detector}

\begin{figure}
\begin{center}
\includegraphics[width=4.in]{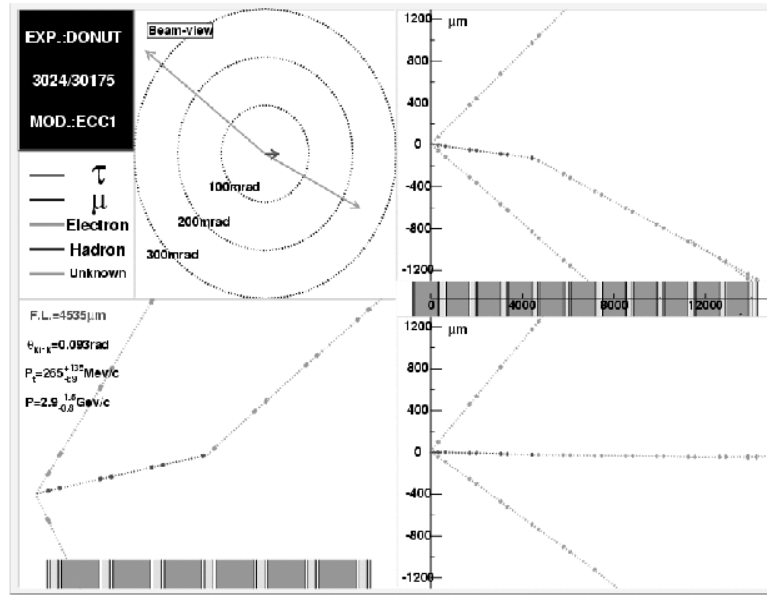}
\end{center}
\caption{Schematic view of a $\tau$ decay candidate observed in the
ECC of the DONUT experiment.}
\label{donut}
\end{figure}

The Emulsion Cloud Chamber (ECC) concept (see references quoted in
\cite{operaproposal}) combines the high-precision tracking
capabilities of nuclear emulsions and the large mass achievable by
employing metal plates as a target. It has been adopted by the OPERA
Collaboration~\cite{operaproposal} for a long-baseline search for
$\nu_{\mu}\rightarrow\nu_{\tau}$ oscillations in the CNGS beam through
the direct detection of the decay of the $\tau$ produced in
$\nu_{\tau}$ CC interactions. As an example of $\tau$ detection, we
show in Fig.~\ref{donut} one of the $\nu_\tau$ events observed in the
DONUT experiment~\cite{Kodama:2000mp}.

The basic element of the OPERA ECC is a ``cell'' made of a 1~mm thick
lead plate followed by a thin emulsion film. The film consists of
$44~\mu$m-thick emulsion layers on either side of a $200~\mu$m plastic
base. The number of grains hits in each emulsion layer ($15$-$20$)
ensures redundancy in the measurement of particle trajectories and
allows the measurement of their energy loss that, in the
non-relativistic regime, can help to distinguish between different
mass hypotheses.

Thanks to the dense ECC structure and to the high granularity provided
by the nuclear emulsions, the OPERA detector is also suited for
electron and $\gamma$ detection.  The resolution in measuring the
energy of an electromagnetic shower is about 20\%. Nuclear emulsions
are able to measure the number of grains associated to each
track. This allows an excellent two-track separation: ${\cal
O}(1~\mu$m) or even better. Therefore, it is possible to disentangle
single-electron tracks from electron pairs coming from $\gamma$
conversion in the lead.

The outstanding position resolution of nuclear emulsions can also be
used to measure the angle of each track segment with an accuracy of
about 1~mrad. This allows a momentum measurement by using multiple
Coulomb scattering, with a resolution of about 20\%, and the
reconstruction of kinematical variables characterizing the event.

A lead-emulsion detector has been also proposed to operate at a Neutrino
Factory to study the so-called silver channel
$\nu_e\rightarrow\nu_{\tau}$~\cite{donini}.

\subsection{Liquid Argon Time Projection Chamber}
The ICARUS Collaboration~\cite{Aprili:2002wx} has developed a liquid argon
time projection chamber (TPC). It is continuously sensitive and
self-triggering, with the ability to provide three-dimensional imaging
of ionizing tracks. 

The operating principle is rather simple: any ionizing event (from a
particle decay or interaction) taking place in the active liquid argon
volume, which is maintained at a temperature $T \sim 89$~K, produces
ion-electron pairs. In the presence of a strong electric field ($\sim
0.5$~KeV/cm), the ions and electrons drift. The motion of the faster
electrons induces a current on a wire plane located near the end of
the sensitive volume. The electrons are collected by a wire plane with
a different orientation. The knowledge of the wire positions and the
drift time provides the three-dimensional image of the track, while
the charge collected on the wires provides information on the
deposited energy.

The detector consists of a large vessel of liquid
argon filled with planes of wires strung on the different orientations. This
device allows tracking, $dE/dx$
measurements and a full-sampling electromagnetic and hadronic
calorimetry. Furthermore, the imaging provides excellent electron and
photon identification and electron/hadron separation. The energy
resolution is excellent for electromagnetic showers ($\sim
13\%/\sqrt{E\mbox{(MeV)}}$) and also very good for contained hadronic
showers ($30\%/\sqrt{E\mbox{(GeV)}}$). Furthermore, it is possible to
measure the momentum of stopping muons with a resolution better than
3\%, by using the Multiple Coulomb Scattering.

\begin{figure}
\begin{center}
\includegraphics[width=4.in]{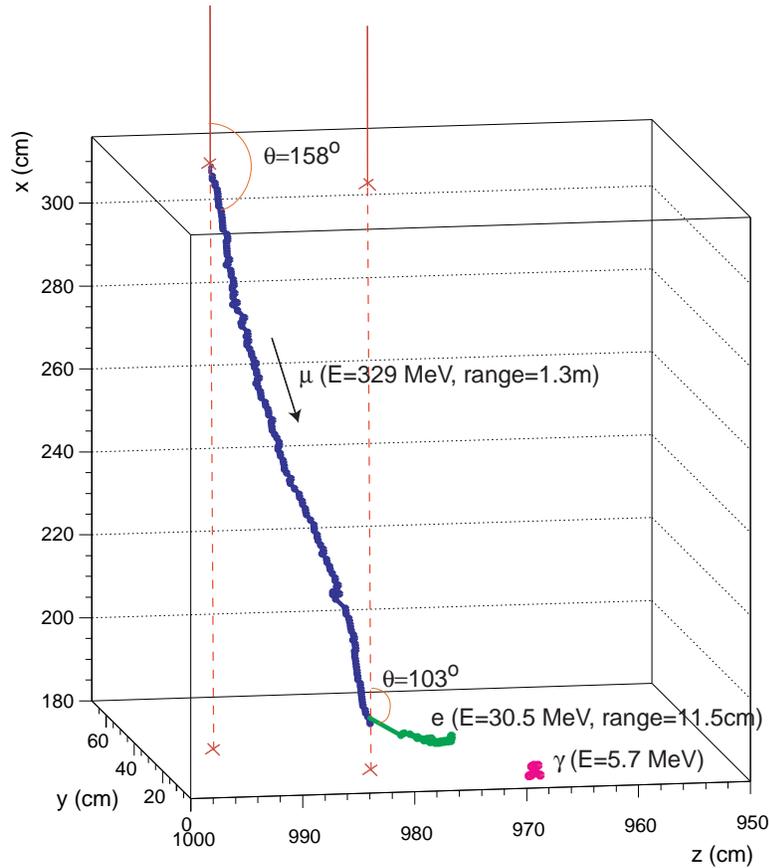}
\end{center}
\caption{Stopping muon in the ICARUS 600~tons detector and decaying
into an electron.}
\label{icarus_eve}
\end{figure}

The major milestone of this technique has been the successful
operation of the ICARUS 600~tons prototype which has operated during
the summer of 2001. An event recorded with the 600~tons detector is shown
in Fig.~\ref{icarus_eve}. Of course neutrino oscillation studies
require an increase of the total liquid argon mass. A 2.5~kton
detector is foreseen to operate with the CNGS neutrino beam to search
for $\nu_\mu\rightarrow\nu_\tau$ oscillations at the atmospheric
scale. Given its excellent electron identification capabilities, it has
been also proposed to operate on other beams to search for
$\nu_\mu\rightarrow\nu_e$ appearance.

\subsection{Low Z Calorimeter}

Unlike the iron calorimeters discussed in Section~\ref{irondet}, low Z
calorimeters allow a good identification and energy measurement of
electrons produced in $\nu_e$CC interactions. In fact, for this
purpose one must sample showers more frequently than about 1.4~X$_0$
and a magnetic field is not necessary. Another advantage of a low Z
calorimeter is that, for a given sampling in units of radiation
length, one can have up to a factor 3 more mass per readout plane with
respect to iron calorimeters. This detector can discriminate between
NC and $\nu_e$ induced CC events by looking at the longitudinal
profile of the neutrino interaction, as NC events are likely to be
much more spread out in the detector than $\nu_e$CC. Several active
detectors (Resistive Plate Chambers, streamer tubes, plastic
scintillators) have been considered and are currently under
investigation. This technique has been proposed to search for
$\nu_{\mu}\rightarrow\nu_e$ oscillations in the NUMI Off-Axis
beam-line ~\cite{Ayres:2002nm}.

\section{Physics Reach of Future Accelerator-Based Oscillation Experiments}

\subsection{Near Term Programs}
The NuMI (expected to start beginning 2005) and the CNGS (expected to
start mid 2006) programs were approved with the aim to search for
neutrino oscillation in the $\Delta m^2$ region indicated by the
atmospheric neutrino results. By looking at the $\nu_\mu$
disappearance in an almost pure $\nu_\mu$ beam, the MINOS
experiment~\cite{MINOS2} at NuMI aims at the measurement of the
oscillation parameters with a precision of about 10\%. A statistical
evidence of $\nu_{\mu}\rightarrow\nu_\tau$ oscillations is achievable
by looking at the NC/CC ratio. On the other hand, the CNGS program
aims at the direct evidence of $\nu_\tau$ appearance in a pure
$\nu_\mu$ beam by searching for the decays of $\tau$ produced in
$\nu_\tau$ CC interactions. After five years of data taking and with
$\Delta m^2_{23} = 2\times10^{-3}~\mbox{eV}^2$, about 10 events, with
a background smaller than 1 event, are expected both in ICARUS and
OPERA~\cite{Duchesneau:2002yq}. A 50\% increase of the proton
intensity ~\cite{Cappi:2001au}, with a corresponding increase of the
event rate, is envisaged. These experiments are also able to search
for $\nu_{\mu}\rightarrow\nu_e$ oscillations, i.e. a non vanishing
$\theta_{13}$. It has been shown in
Ref.~\cite{Migliozzi:2003pw,migliozzi} that in case of negative
result, the CNGS program (ICARUS and OPERA together) will be able to
exclude values of $\theta_{13}$ down to $5^\circ$ at 90\%~C.L., to be
compared with the $\theta_{13}< 10^\circ$ limit given by CHOOZ. On the
other hand, if $\theta_{13}$ is larger than $7^\circ$ it will be
possible to provide a first measurement of $\theta_{13}$.

\subsection{Medium Term Programs}
The main goals of the J-PARC~\cite{Itow:2001ee} (expected to start in
2008-09) and NuMI-OA~\cite{Ayres:2002nm} (start of data taking after
2010) programs are to provide precision measurements of the PMNS
matrix elements relevant in the atmospheric sector with a few percent
accuracy and to push the sensitivity to $\theta_{13}$ down to a few
degrees, i.e. improve the CHOOZ limit by one order of magnitude. As
far as the detector is concerned, the J-PARC project will exploit the
Super-Kamiokande detector located about 300~km far from the neutrino
source under construction at the Tokai site in Japan. The
NuMI-OA project will exploit the NuMI beam with a low Z calorimeter
located at a distance not yet defined, but in the 700 - 900~km range.

It is worth to stress that the information gathered from the
combination of results from different experiments is actually better
than if one had simply run a single experiment
longer~\cite{Huber:2002rs}. The reason for this is that NuMI-OA
experiment can be sensitive to matter effects, while J-PARC experiment
is not.

\subsection{Long Term Programs}
\label{far}
The main goal of future projects like Super-Beams
(i.e. J-PARC-II~\cite{prism}, Super AGS~\cite{Diwan:2003bp},
SPL~\cite{spl,Apollonio:2002en}), $\beta$-beams or Neutrino Factories
is the discovery and the measurement of the CP violating phase
$\delta$ in the leptonic sector. This search is very challenging and
intrinsically complex. As an example, we report about the search for
$\delta$ at a Neutrino Factory.

The transition probabilities $\nu_e \to \nu_\mu$ and $\bar{\nu}_e \to
\bar{\nu}_\mu$ are extremely sensitive to $\theta_{13}$ and $\delta$:
these are the so-called ``golden measurements''~\cite{Cervera:2000kp}
and can be studied by searching for wrong-sign muons, provided the
considered detector has a good muon charge identification
capability. The determination of ($\theta_{13},\delta$) from this
channel is not free of ambiguities: it was shown
in~\cite{Burguet-Castell:2001ez} that, for a given physical input
parameter pair ($\bar \theta_{13},\bar \delta$), measurements of the
oscillation probability for $\nu_e \to \nu_\mu$ and $\bar \nu_e \to
\bar \nu_\mu$ generally result in two allowed regions of the parameter
space. Worse than that, further degeneracies results from our
ignorance of the sign of $\Delta m^2_{23}$ and from the approximate
$[\theta_{23}, \pi/2 - \theta_{23}]$ symmetry for the atmospheric
angle~\cite{Minakata:2001qm,Barger:2001yr}. In general, the
measurement of $P(\nu_e \to \nu_\mu)$ and $P(\bar \nu_e \to \bar
\nu_\mu)$ will result in eight allowed regions of the parameter space,
the so-called eightfold-degeneracy~\cite{Barger:2001yr}.

\begin{figure}
\begin{center}
\includegraphics[width=2in]{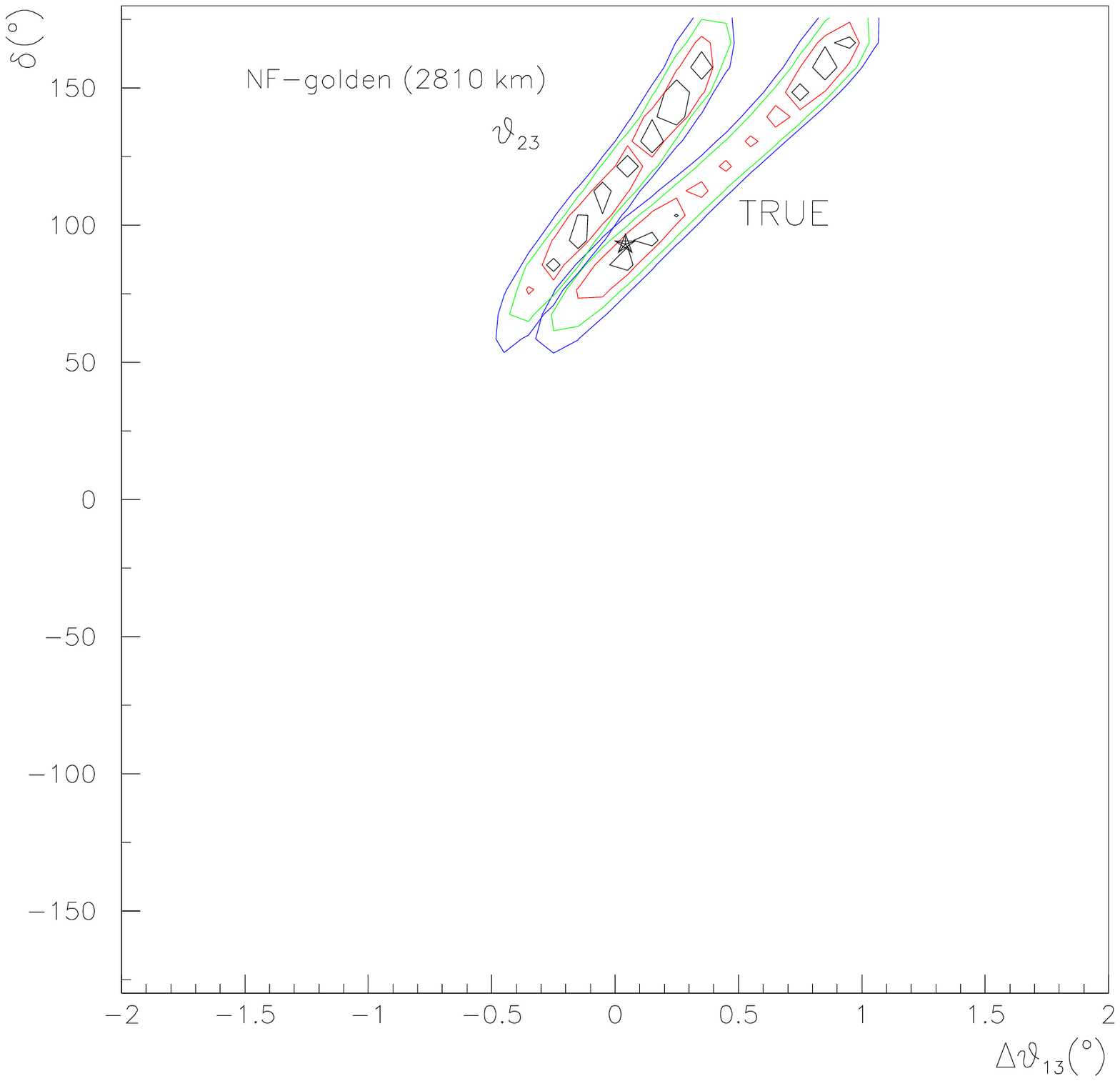} \includegraphics[width=2in]{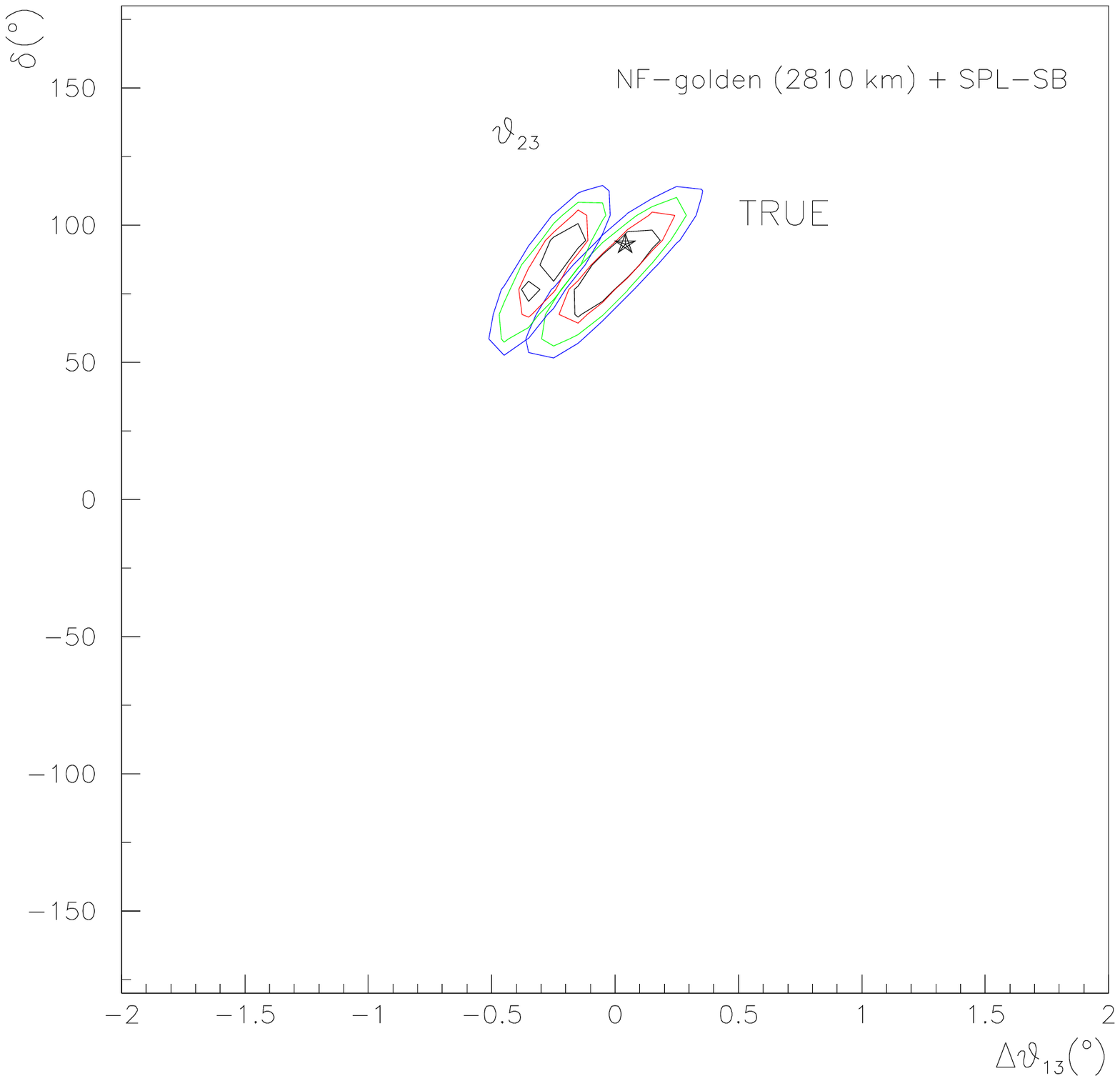} \\
\includegraphics[width=2in]{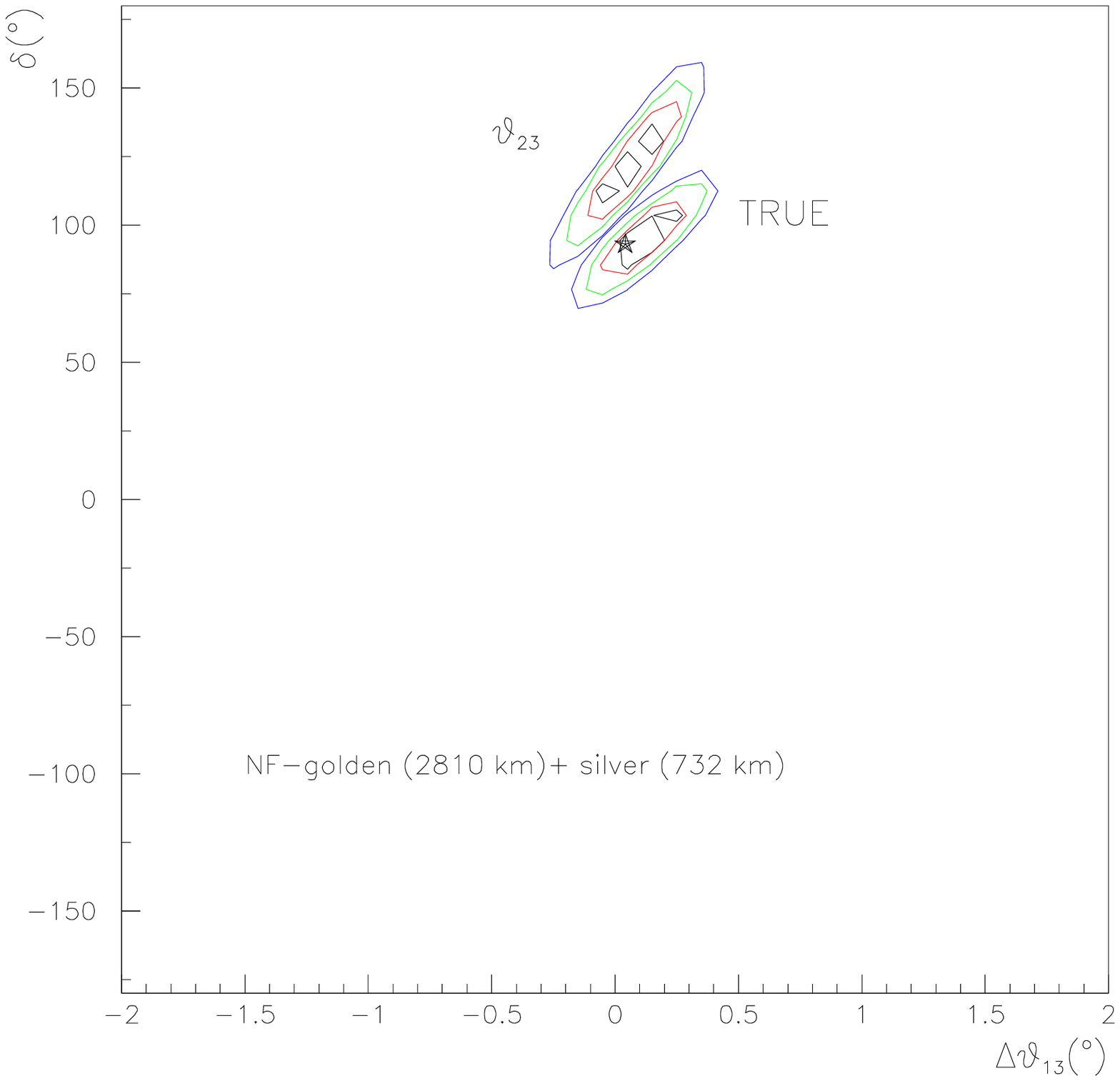} \includegraphics[width=2in]{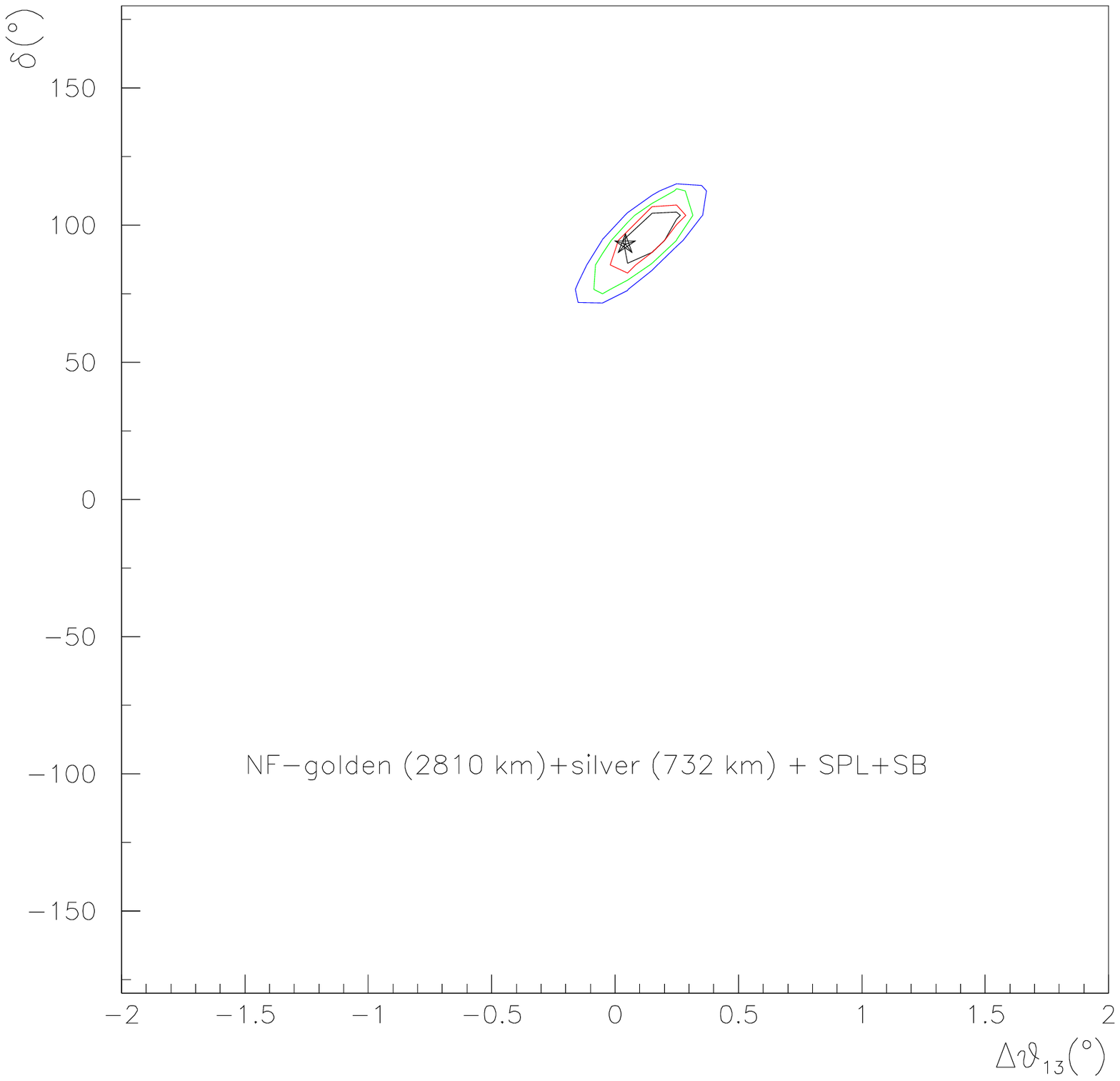}
\end{center}
    \caption{The results of a $\chi^2$ fit for $\bar \theta_{13} =
2^\circ; \bar \delta = 90^ \circ$.  Four different combinations of
experimental data are presented: a) magnetized iron detector (MID) at
a Neutrino Factory (NF); b) MID at NF plus water Cerenkov (WC) at a
SPL; c) MID plus hybrid emulsion (HE) at NF; d) the three detectors
together.  Notice how in case (d) the eightfold-degeneracy is solved
and a good reconstruction of the physical $\theta_{13},\delta$ values
is achieved.}
    \label{super}
\end{figure}

In order to solve these ambiguities, a single experiment on a single
neutrino beam is not enough. An optimal combination of $\beta$-beams,
Super-Beams and Neutrino Factories has to be considered to deal with
the eightfold degeneracy. Several investigations on how to solve this
problem have been carried out, as reported in~\cite{Donini:2003kr} and
references therein. As an example we report in Fig.~\ref{super} (taken
from~\cite{Donini:2003kr}) the result of an analysis which combines a
magnetized iron detector and a hybrid emulsion detector studying at a
Neutrino Factory the golden and the silver channels, respectively, and
a water Cerenkov detector exposed to a Super-Beam. For comparison,
results achievable by using a combination of only two detectors or by
the magnetized iron detector alone are also given.

\subsection{The Implications of Present and Medium Term on the Long Term Programs}

Having illustrated the experimental programs toward a precision
measurement of the PMNS matrix, some comments are in order (for more
details we refer to~\cite{migliozzi}). One would like to approve and
fund on firm basis expensive long term programs. The recent KAMLAND
result strongly supports a high value of $\Delta m_{12}^2$ and a
large, although not maximal, mixing angle. Therefore, this result
places the long term program on a firmer ground since it guarantees
that sub-dominant effects will not be suppressed to an unobservable
level ($\alpha\equiv \Delta m_{12}^2/\Delta m_{23}^2\ll
10^{-2}$). However, there is a second condition, which at present
remains poorly constrained, to ensure that $\delta$ is observable:
$\theta_{13}$ should not be vanishingly small.

\begin{figure}
\begin{center}
\includegraphics[width=4.in]{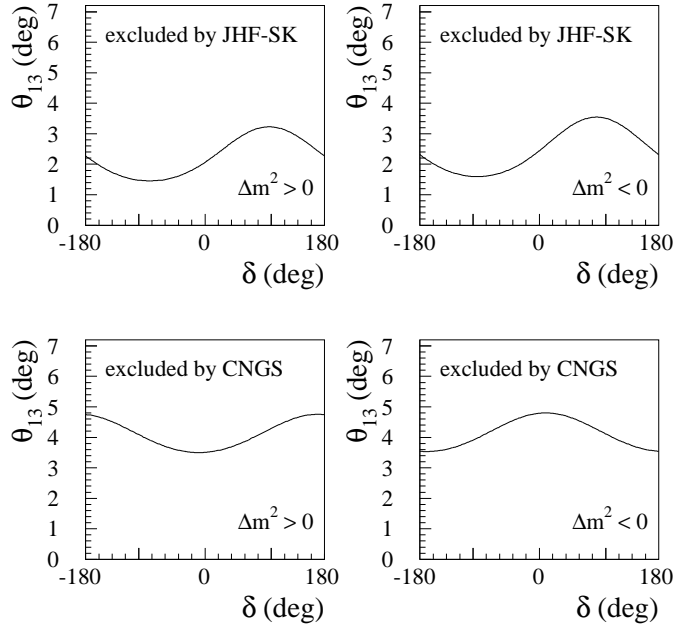}
\end{center}
\caption{$\sin^2 2\theta_{13}$ sensitivity at 90\% C.L. versus $\delta$ and for both positive and negative $\Delta m^2_{23}$. JHK-SK stands for J-PARC.}
\label{excl_2d}
\end{figure}

We then expect present and medium term programs to measure
$\theta_{13}$ in order to assess the possibility of exploring CP
violation through Super-Beams (i.e. J-PARC-II, Super AGS, SPL),
$\beta$-beams or Neutrino Factories. However, effects driven by
$\theta_{13}$ are correlated with the ones driven by $\delta$. As an
example we show in Fig.~\ref{excl_2d} the $\theta_{13}$ sensitivity at
90\% C.L. as a function of $\delta$ for J-PARC and for
CNGS~\cite{migliozzi}. For positive values of $\delta$, the CP phase
dependence of J-PARC has the worst possible behavior for the medium
term program, since the minimum sensitivity to $\sin^22\theta_{13}$ is
achieved at maximum ($90^\circ$) CP violation phase (maximum discovery
potential for long term programs).

The evaluation of the physics reach of the long term experiments will
be particularly difficult if none of the medium term experiments finds
a conclusive $\theta_{13}$ signal. It will be impossible to lift the
$\theta_{13}-\delta$ correlation without additional external
information; i.e. it is impossible to decide whether the lack of
events is due to the smallness of $\theta_{13}$ or it is the outcome
of a cancellation effect between a large value of $\theta_{13}$ and a
large value of $\delta$. Additional ambiguities are present when
matter effects are appreciable. A careful optimization of the whole
medium term program can considerably improve the sensitivity to
$\theta_{13}$.  Therefore, one should optimize the data taking in
order to decouple the $\theta_{13}-\delta$ correlation. In particular,
one could decide to run one of the neutrino beam-lines with
anti-neutrinos.

\begin{figure}
\begin{center}
\includegraphics[width=4.in]{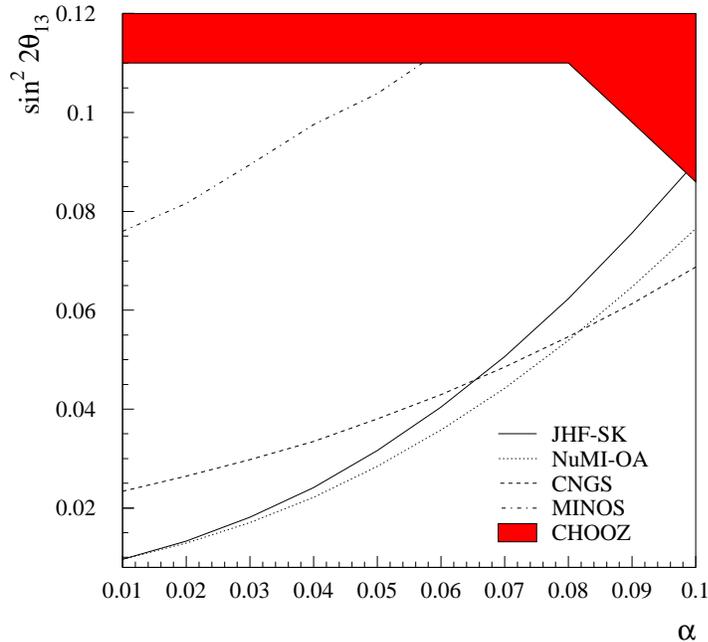}
\end{center}
\caption{$\sin^2 2\theta_{13}$ sensitivity at 90\% C.L. versus $\alpha
\equiv \Delta m^2_{21}/ | \Delta m^2_{31} |$. JHK-SK stands for J-PARC.}
\label{alphadip}
\end{figure}

The $\theta_{13}$ sensitivity achievable by each medium term
experiement (and incorporating all the effect of degeneracies) is
shown in Fig.~\ref{alphadip}~\cite{migliozzi}.

We can conclude that without an optimal running of all available
beam-lines it will be impossible for present and medium term programs
to ground (or discourage in a definitive manner) the building of long
term facilities.

\section{Conclusion}
So far, neutrinos are the only source of physics beyond the Standard
Model of particle physics: the neutrino oscillations discovery has
unambiguously given evidence for a non-zero neutrino mass and shown
the non-conservation of the leptonic number. However, there is a still
long way to go before the mixing matrix in the leptonic sector and the
masses are precisely determined. There are ongoing efforts to provide
new neutrino sources (neutrino beams) and new detector technologies or
upgrades of the existing ones. An overview of all existing, under
construction and proposed beam-lines and detectors is given in
Table~\ref{tab:beams}.

Given these premises, one can think that in the future the year 1998
will remembered not only as the year of the neutrino oscillations
discovery, but as the starting date of a new "mirabilis era" for
neutrino physics.

\begin{table}[tbp]
\begin{center}
{\small
\begin{tabular}{|c|c|c|c|c|}
\hline Program & Detector & $<E>$ & L & evts/kton/yr \\ 
& & (GeV) & (km) & \\ 
\hline 
NuMI & MI & 3.5 & 730 & 469 \\ 
\hline 
CNGS & HE LAr & 18 & 732 & 2500 \\ 
\hline \hline 
J-PARC & WC & 0.7 & 295 & 133 \\
\hline 
NuMI-OA & LZ & 2 & 700 - 900 & 80 \\ 
\hline \hline 
J-PARC-II & WC & 0.7 & 295 & 691 \\ 
\hline 
Super AGS & WC, LAr, LZ & 1.5 & 2540 & 11 \\
\hline 
SPL & WC, LAr & 0.26 & 130 & 16 \\ 
\hline 
$\beta$-beam & WC & 0.58 & 130 & 84 \\ 
\hline 
$\nu$-Factory & MI, HE, LAr & 10-30 & 700-3000 & $10^5-10^3$ \\ 
\hline

\end{tabular}
}
\end{center}
\caption{Main features of future long base-line neutrino oscillation
experiments based on accelerators and the proposed detector
techniques. In the Table we used the following conventions:
WC$\equiv$Water Cerenkov; MI$\equiv$Magnetized Iron; HE$\equiv$Hybrid
Emulsion; LAr$\equiv$Liquid Argon; LZ$\equiv$Low Z.}
\label{tab:beams}

\end{table}

%%%%%%%%%%%%%%%%%%%%%%%%%%%%%%%%%%%%%%%%%%%%%%%%%%%%%%%%%%%%%%%%%%%%%%%%%%%%% 
% Bibliography
%%%%%%%%%%%%%%%%%%%%%%%%%%%%%%%%%%%%%%%%%%%%%%%%%%%%%%%%%%%%%%%%%%%%%%%%%%%%%%
%


\baselineskip=10pt \vspace{1cm}
\begin{thebibliography}{99}

%\cite{Pontecorvo:1957yb}
\bibitem{Pontecorvo:1957yb}
B.~Pontecorvo,
%``Mesonium And Antimesonium,''
Sov.\ Phys.\ JETP {\bf 6} (1957) 429
[Zh.\ Eksp.\ Teor.\ Fiz.\  {\bf 33} (1957) 549];\\
%%CITATION = SPHJA,6,429.1957\ ZETFA,33,549;%%
Z.~Maki, M.~Nakagawa and S.~Sakata,
%``Remarks On The Unified Model Of Elementary Particles,''
Prog.\ Theor.\ Phys.\ {\bf 28} (1962) 870;\\
%%CITATION = PTPKA,28,870;%%
B.~Pontecorvo,
%``Neutrino experiments and the question of leptonic-charge  conservation,''
Sov.\ Phys.\ JETP {\bf 26} (1968) 984;\\
%%CITATION = SPHJA,26,984;%%
V.~N.~Gribov and B.~Pontecorvo,
%``Neutrino Astronomy And Lepton Charge,''
Phys.\ Lett.\ B {\bf 28} (1969) 493.
%%CITATION = PHLTA,B28,493;%%
\bibitem{atmo}
Y.~Fukuda {\it et al.}  [Super-Kamiokande Coll.],
Phys.\ Rev.\ Lett.\  {\bf 81}, 1562 (1998); \\
M.~Ambrosio {\it et al.}  [MACRO Coll.],
Phys.\ Lett.\ {\bf B517} (2001) 59.
\bibitem{solar}
B.T.~Cleveland {\it et al.},
Astrophys.\ J.\  {\bf 496}, 505 (1998); \\
J.N.~Abdurashitov {\it et al.}  [SAGE Coll.],
Phys.\ Rev.\ {\bf C60}, 055801 (1999); \\
W.~Hampel {\it et al.}  [GALLEX Coll.],
Phys.\ Lett.\ {\bf B447}, 127 (1999); \\
S.~Fukuda {\it et al.}  [Super-Kamiokande Coll.],
Phys.\ Rev.\ Lett.\  {\bf 86}, 5651 (2001); \\
Q.R.~Ahmad {\it et al.}  [SNO Coll.],
Phys.\ Rev.\ Lett.\  {\bf 87}, 071301 (2001).

\bibitem{k2k}
M.H.~Ahn {\it et al.}  [K2K Coll.],
Phys.\ Rev.\ Lett.\  {\bf 90} (2003) 041801.

\bibitem{KAMLAND}
K.~Eguchi {\it et al.}  [KamLAND Coll.],
Phys.\ Rev.\ Lett.\  {\bf 90} (2003) 021802.

%\cite{Athanassopoulos:1997pv}
\bibitem{Athanassopoulos:1998pv}
C.~Athanassopoulos {\it et al.}  [LSND Collaboration],
%``Evidence for nu/mu $\to$ nu/e neutrino oscillations from LSND,''
Phys.\ Rev.\ Lett.\  {\bf 81} (1998) 1774
[arXiv:nucl-ex/9709006];
%%CITATION = NUCL-EX 9709006;%%
A.~Aguilar {\it et al.}  [LSND Collaboration],
%``Evidence for neutrino oscillations from the observation of anti-nu/e  appearance in a anti-nu/mu beam,''
Phys.\ Rev.\ D {\bf 64} (2001) 112007
[arXiv:hep-ex/0104049].
%%CITATION = HEP-EX 0104049;%%

\bibitem{miniboone} Details on the MiniBooNE detector are available at http://www-boone.fnal.gov/

%\cite{Maltoni:2003da}
\bibitem{Maltoni:2003da}
M.~Maltoni, T.~Schwetz, M.~A.~Tortola and J.~W.~Valle,
%``Status of three-neutrino oscillations after the SNO-salt data,''
arXiv:hep-ph/0309130.
%%CITATION = HEP-PH 0309130;%%

%\cite{Apollonio:2002gd}
\bibitem{Apollonio:2002gd}
M.~Apollonio {\it et al.}  [CHOOZ Collaboration],
%``Limits on neutrino oscillations from the CHOOZ experiment,''
Phys.\ Lett.\ B {\bf 466} (1999) 415
[arXiv:hep-ex/9907037];
%%CITATION = HEP-EX 9907037;%%

M.~Apollonio {\it et al.},
%``Search for neutrino oscillations on a long base-line at the CHOOZ  nuclear power station,''
Eur.\ Phys.\ J.\ C {\bf 27} (2003) 331
[arXiv:hep-ex/0301017].
%%CITATION = HEP-EX 0301017;%%

%\cite{Alberico:2003kd}
\bibitem{review}
W.~M.~Alberico and S.~M.~Bilenky,
%``Neutrino oscillations, masses and mixing,''
arXiv:hep-ph/0306239.
%%CITATION = HEP-PH 0306239;%%

S.~M.~Bilenky, C.~Giunti, J.~A.~Grifols and E.~Masso,
%``Absolute values of neutrino masses: Status and prospects,''
Phys.\ Rept.\  {\bf 379} (2003) 69
[arXiv:hep-ph/0211462].
%%CITATION = HEP-PH 0211462;%%

%\cite{Apollonio:2002en}
\bibitem{Apollonio:2002en}
M.~Apollonio {\it et al.},
%``Oscillation physics with a neutrino factory. ((G)) ((U)),''
arXiv:hep-ph/0210192.
%%CITATION = HEP-PH 0210192;%%

\bibitem{beavis}
D.~Beavis {\it et al.}, BNL No. 52459, April 1995.

\bibitem{kaji} Courtesy of T.~Kajita.

%\cite{Kobayashi:sc}
\bibitem{Kobayashi:sc}
T.~Kobayashi,
%``The Jhf Neutrino Experiment,''
Nucl.\ Phys.\ Proc.\ Suppl.\  {\bf 111} (2002) 163.
%%CITATION = NUPHZ,111,163;%%

%\cite{Zucchelli:sa}
\bibitem{Zucchelli:sa}
P.~Zucchelli,
%``A Novel Concept For A Anti-Nu/E / Nu/E Neutrino Factory: The Beta Beam,''
Phys.\ Lett.\ B {\bf 532} (2002) 166.
%%CITATION = PHLTA,B532,166;%%

%\cite{donini}
\bibitem{donini}
A.~Donini, D.~Meloni and P.~Migliozzi,
%``The silver channel at the neutrino factory,''
Nucl.\ Phys.\ B {\bf 646} (2002) 321
[arXiv:hep-ph/0206034].
%%CITATION = HEP-PH 0206034;%%

D.~Autiero {\it et al.},
%``The synergy of the golden and silver channels at the Neutrino Factory,''
arXiv:hep-ph/0305185.
%%CITATION = HEP-PH 0305185;%%

%\cite{Harris:2003rc}
\bibitem{Harris:2003rc}
D.~A.~Harris,
%``Future experiments with neutrino superbeams, beta-beams and neutrino factories,''
FERMILAB-CONF-03-328-E
%\href{http://www.slac.stanford.edu/spires/find/hep/www?r=fermilab-conf-03-328-e}{SPIRES entry}

%\cite{Boger:1999bb}
\bibitem{Boger:1999bb}
J.~Boger {\it et al.}  [SNO Collaboration],
%``The Sudbury Neutrino Observatory,''
Nucl.\ Instrum.\ Meth.\ A {\bf 449} (2000) 172
[arXiv:nucl-ex/9910016].
%%CITATION = NUCL-EX 9910016;%%

%\cite{Ayres:2002nm}
\bibitem{minos}
M.~V.~Diwan,  "Invited talk at the Seventh International Workshop on Tau Lepton Physics (TAU02), Santa Cruz, Ca, USA, Sept 2002",
%``Status of three-neutrino oscillations after the SNO-salt data,''
arXiv:hep-ex/0211026.

%\cite{Cervera:2000kp}
\bibitem{Cervera:2000kp}
A.~Cervera, A.~Donini, M.~B.~Gavela, J.~J.~Gomez Cadenas, P.~Hernandez, O.~Mena and S.~Rigolin,
%``Golden measurements at a neutrino factory,''
Nucl.\ Phys.\ B {\bf 579} (2000) 17
[Erratum-ibid.\ B {\bf 593} (2001) 731]
[arXiv:hep-ph/0002108].
%%CITATION = HEP-PH 0002108;%%

%\cite{Guler:2000bd}
\bibitem{operaproposal}
M.~Guler {\it et al.}  [OPERA Collaboration],
%``OPERA: An appearance experiment to search for nu/mu <--> nu/tau  oscillations in the CNGS beam. Experimental proposal,''
CERN-SPSC-2000-028;
%\href{http://www.slac.stanford.edu/spires/find/hep/www?r=cern-spsc-2000-028}{SPIRES entry}

K.~Kodama {\it et al.}  [OPERA Collaboration],
%``A long baseline nu/tau appearance experiment in the CNGS beam from  CERN to Gran Sasso. Progress report,''
CERN-SPSC-99-20.
%\href{http://www.slac.stanford.edu/spires/find/hep/www?r=cern-spsc-99-20}{SPIRES entry}

%\cite{Kodama:2000mp}
\bibitem{Kodama:2000mp}
K.~Kodama {\it et al.}  [DONUT Collaboration],
%``Observation of tau-neutrino interactions,''
Phys.\ Lett.\ B {\bf 504} (2001) 218
[arXiv:hep-ex/0012035].
%%CITATION = HEP-EX 0012035;%%
K.~Kodama {\it et al.},
%``Detection And Analysis Of Tau Neutrino Interactions In Donut Emulsion Target,''
Nucl.\ Instrum.\ Meth.\ A {\bf 493} (2002) 45.
%%CITATION = NUIMA,A493,45;%%

%\cite{Aprili:2002wx}
\bibitem{Aprili:2002wx}
P.~Aprili {\it et al.}  [ICARUS Collaboration],
%``The ICARUS experiment: A second-generation proton decay experiment and neutrino observatory at the Gran Sasso laboratory. Cloning of T600
modules to reach the design sensitive mass. (Addendum),''
CERN-SPSC-2002-027
%\href{http://www.slac.stanford.edu/spires/find/hep/www?r=cern-spsc-2002-027}{SPIRES entry}

\bibitem{MINOS2}
MINOS Technical Design Report NuMI-L-337 TDR,\\
{\tt http://www-numi.fnal.gov:8875/minwork/info/tdr/}, \\
{\tt http://www-numi.fnal.gov:8875/forscientists.html}; \\
V. Paolone, Nucl. Phys. Proc. Suppl. {\bf 100} (2001) 197;
S.~G.~Wojcicki,
%``Status Of The Minos Experiment,''
Nucl.\ Phys.\ Proc.\ Suppl.\  {\bf 91} (2001) 216.

%\cite{Duchesneau:2002yq}
\bibitem{Duchesneau:2002yq}
D.~Duchesneau  [OPERA Collaboration],
%``The CERN - Gran Sasso neutrino program,''
eConf {\bf C0209101} (2002) TH09
[arXiv:hep-ex/0209082].
%%CITATION = HEP-EX 0209082;%%

%\cite{Itow:2001ee}
\bibitem{Itow:2001ee}
Y.~Itow {\it et al.},
%``The JHF-Kamioka neutrino project,''
arXiv:hep-ex/0106019.
%%CITATION = HEP-EX 0106019;%%

\bibitem{Ayres:2002nm}
D.~Ayres {\it et al.},
%``Letter of intent to build an off-axis detector to study nu/mu $\to$ nu/e oscillations with the NuMI neutrino beam,''
arXiv:hep-ex/0210005.
%%CITATION = HEP-EX 0210005;%%

%\cite{Cappi:2001au}
\bibitem{Cappi:2001au}
R.~Cappi {\it et al.},
%``Increasing the Proton Intensity of PS and SPS,''
CERN-PS-2001-041-AE.
%\href{http://www.slac.stanford.edu/spires/find/hep/www?r=cern-ps-2001-041-ae}{SPIRES entry}


%\cite{Migliozzi:2003pw}
\bibitem{Migliozzi:2003pw}
M.~Komatsu, P.~Migliozzi and F.~Terranova,
%``Sensitivity to Theta(13) of the CERN to Gran Sasso neutrino beam,''
J.\ Phys.\ G {\bf 29} (2003) 443
[arXiv:hep-ph/0210043];
%%CITATION = HEP-PH 0210043;%%

\bibitem{migliozzi}
P.~Migliozzi and F.~Terranova,
%``Next generation long baseline experiments on the path to leptonic CP  violation,''
Phys.\ Lett.\ B {\bf 563} (2003) 73
[arXiv:hep-ph/0302274].
%%CITATION = HEP-PH 0302274;%%

%\cite{Huber:2002rs}
\bibitem{Huber:2002rs}
P.~Huber, M.~Lindner and W.~Winter,
%``Synergies between the first-generation JHF-SK and NuMI superbeam experiments,''
Nucl.\ Phys.\ B {\bf 654} (2003) 3
[arXiv:hep-ph/0211300].
%%CITATION = HEP-PH 0211300;%%

\bibitem{prism}
M. Furusaka {\it et al.}, JAERI/KEK Joint Project Proposal 
KEK-REPORT-99-4, JAERI-TECH-99-01.

%\cite{Diwan:2003bp}
\bibitem{Diwan:2003bp}
M.~V.~Diwan {\it et al.},
%``Very long baseline neutrino oscillation experiments for precise  measurements of mixing parameters and CP violating effects,''
Phys.\ Rev.\ D {\bf 68} (2003) 012002
[arXiv:hep-ph/0303081].
%%CITATION = HEP-PH 0303081;%%

\bibitem{spl}
B. Autin {\it et al.}, CERN report CERN-2000-012 (2000).

%\cite{Burguet-Castell:2001ez}
\bibitem{Burguet-Castell:2001ez}
J.~Burguet-Castell, M.~B.~Gavela, J.~J.~Gomez-Cadenas, P.~Hernandez and O.~Mena,
%``On the measurement of leptonic CP violation,''
Nucl.\ Phys.\ B {\bf 608} (2001) 301
[arXiv:hep-ph/0103258].
%%CITATION = HEP-PH 0103258;%%

%\cite{Minakata:2001qm}
\bibitem{Minakata:2001qm}
H.~Minakata and H.~Nunokawa,
%``Exploring neutrino mixing with low energy superbeams,''
JHEP {\bf 0110} (2001) 001
[arXiv:hep-ph/0108085].
%%CITATION = HEP-PH 0108085;%%

%\cite{Barger:2001yr}
\bibitem{Barger:2001yr}
V.~Barger, D.~Marfatia and K.~Whisnant,
%``Breaking eight-fold degeneracies in neutrino CP violation, mixing, and  mass hierarchy,''
Phys.\ Rev.\ D {\bf 65} (2002) 073023
[arXiv:hep-ph/0112119].
%%CITATION = HEP-PH 0112119;%%

%\cite{Donini:2003kr}
\bibitem{Donini:2003kr}
A.~Donini,
``NUFACT'03: The fate of the clones,''
arXiv:hep-ph/0310014.
%%CITATION = HEP-PH 0310014;%%


\end{thebibliography}
\end{document}